\documentclass[a4paper,10pt]{article}

\newcommand{\bfa}[1]{\mbox{\boldmath${#1}$}}

\newcommand{\bea}{\begin{eqnarray}}
\newcommand{\eea}{\end{eqnarray}}
\newcommand{\bnn}{\begin{eqnarray*}}
\newcommand{\enn}{\end{eqnarray*}}
\newcommand{\be}{\begin{equation}}
\newcommand{\ee}{\end{equation}}

\def\PACS{\par\leavevmode\hbox {\it PACS:\ }}%
\def\MSC{\par\leavevmode\hbox {\it MSC:\ }}%
\def\UK{\par\leavevmode\hbox {\it Keywords:\ }}%

\begin{document}

\title{Spinors in the hyperbolic algebra}

\author{S. Ulrych\\Wehrenbachhalde 35, CH-8053 Z\"urich, Switzerland}
\date{October 16, 2005}
\maketitle

\begin{abstract}
The three-dimensional universal complex Clifford algebra
$\bar{\bfa{C}}_{3,0}$ is used to represent relativistic vectors in terms of paravectors.
In analogy to the Hestenes spacetime approach spinors are
introduced in an algebraic form. This removes the dependance on
an explicit matrix representation of the algebra.
\end{abstract}

{\scriptsize\PACS{03.65.Fd; 02.10.De; 11.30.Cp; 02.10.Hh; 02.20.Qs}
\MSC{81R25; 15A66; 30G35; 15A33; 20H25}
\UK{Hyperbolic complex Clifford algebra; Algebraic spinor; Hyperbolic numbers; Paracomplex numbers; Split-complex numbers}}

\section{Introduction}
Over the last years there is growing interest in the Clfford algebra approach
to spacetime that has been initiated by Hestenes \cite{Hes66,Hes84,Hes99,Hes03}.
The algebraic representation of geometry has in general advantages compared to the
conventional description in terms of column vectors and matrices. An overview
of applications is given by Doran and Lasenby \cite{Dor03} and by Gull et al. \cite{Gul93}.

Beside the approach to relativistic physics in terms of the $\bfa{R}_{1,3}$ Dirac algebra,
there is the less noticed approach of Baylis which is based on $\bfa{R}_{3,0}$ paravectors \cite{Bay89}.
Baylis derived a new representation of electrodynamics \cite{Bay99} with these paravectors.
This algebra has been introduced originally by Sobczyk in the spacetime
vector analysis \cite{Sob81}.
In a recent work \cite{Ulr05} this algebra is generalized to the universal
complex Clifford algebra $\bar{\bfa{C}}_{3,0}$. The structural difference compared to Baylis
appears in the shape of the hyperbolic unit, which plays an integral part in the complex formalism.

For more details on the hyperbolic numbers and their properties it is referred to the
references in \cite{Ulr05}. However, it should be mentioned that
hyperbolic numbers are used also within general relativity, where the hyperbolic numbers
are also denoted as paracomplex or split-complex numbers.
Paracomplex projective models and harmonic maps were investigated by Erdem \cite{Erd83,Erd97,Erd99}.
A survey on paracomplex geometry,
para-Hermitian, and para-Kaehler manifolds has been given by Cruceanu et al. \cite{Cru95,Cru96}.
Solutions of Minkowskian sigma models generated by hyperbolic numbers were considered
by Lambert et al. \cite{Lam87,Lam88}. Zhong generated
new solutions of the stationary axisymmetric Einstein equations with hyperbolic numbers \cite{Zho85}.
He investigated hyperbolic complex linear symmetry groups and their local gauge
transformation actions \cite{Zho85a}. Furthermore,
the hyperbolic complexification of Hopf algebras \cite{Zho99}.
Moffat \cite{Mof82} has interpretated the hyperbolic number as fermion number.
This interpretation has led to fundamental explanation
of stability of fermionic matter.

As a supplement to the references in \cite{Ulr05} it is mentioned that
introductions to hyperbolic numbers including further references are given
by Sobczyk \cite{Sob95}, Borota, and Osler \cite{Bor02}. 
Hyperbolic numbers are applied to integrable systems by Bracken and Hayes \cite{Bra02,Bra03}.
Linear and quasilinear complex equations are
investigated by Wen based on hyperbolic numbers \cite{Wen02}.
A slightly different structure than the algebra used in this work have the so-called paraquaternions
(or split-quaternions \cite{Ino04})
used, e.g., by Bla\v zi\'c \cite{Bla96}.

The hyperbolic numbers form, together with complex numbers and quaternions, the fundamental
building blocks in the classification of Clifford algebras. Porteous \cite{Por95} gives an overview
of real and complex Clifford algebras for geometries with arbitrary signatures. Beside the real
Clifford algebras there exist four different types of complex algebras. Porteous derives explicit
matrix representations for all algebras.
Spinors are elements of a minimal left ideal, which can be represented as column vectors
corresponding to the matrix representation of the algebra.
Such a representation has been used in \cite{Ulr05}. 

Though this conventional picture is familiar to physicists, Hestenes actually promoted
the full algebraic representation of vectors and spinors. The theory should
be free of any explicit matrix repesentations. Hestenes achieved this for non-relativistic and
relativistic physics. 
It is the intention of this work to apply these ideas explicitly to the three-dimensional
complex paravector algebra.

\section{Hyperbolic algebra}
\label{hypalg}
Vector spaces can be defined over the commutative ring of 
hyperbolic numbers $z\in\bfa{H}$
\be
\label{beg}
z=x+iy+jv+ijw\;,\hspace{0.5cm}x,y,v,w \in\bfa{R}\;,
\ee
where the hyperbolic unit $j$ has the property $j^2=1$. 
In the terminology of Clifford algebras
the hyperbolic numbers defined in this way are represented by $\bar{\bfa{C}}_{1,0}$, i.e.,
they correspond to the universal one-dimensional complex Clifford algebra (see Porteous \cite{Por95}).

Beside the grade involution, two anti-involutions play a major role in the
description of Clifford algebras and their structure, conjugation and reversion.
Conjugation changes the sign of the complex and
the hyperbolic unit
\be
\label{conj}
\bar{z}=x-iy-jv+ijw\;.
\ee 
Reversion, denoted as $z^\dagger$, changes only the sign of the complex unit.
Anti-involutions reverse the ordering in the multiplication, e.g., $(ab)^\dagger=b^\dagger a^\dagger$.
This becomes
important when non-commuting elements of an algebra are considered.
With respect to conjugation the square of the 
hyperbolic number,
\be
\label{sqsymb}
\vert z\vert^2=z\bar{z}\;,
\ee
can be calculated as
\be
\label{square}
z\bar{z}=x^2+y^2-v^2-w^2+2ij(xw-yv)\;.
\ee

The hyperbolic numbers form the basis of the hyperbolic paravector algebra.
A Minkowski vector $x^\mu=(x^0,x^i)\in\bfa{R}^{\,3,1}$ is represented in terms of the
hyperbolic algebra as
\be
\label{veco}
x=x^\mu e_\mu\;.
\ee
The basis elements $e_\mu=(e_0,e_i)$ include the
unity and the Pauli algebra
multiplied by the hyperbolic
unit~$j$ 
\be
 e_\mu=(1,j\sigma_i)\;.
\ee
The algebra can be complexified with either the hyperbolic or the complex unit.
The full structure is equivalent to the universal three-dimensional complex Clifford
algebra $\bar{\bfa{C}}_{3,0}$. 

The scalar product of two vectors
is defined as
\be
\label{scalar}
x\cdot y
=\frac{1}{2}(x\bar{y}+y\bar{x})=\langle x\bar{y}\rangle_+\;.
\ee 
The wedge product is given as
\be
\label{outer}
x\wedge y
=\frac{1}{2}(x\bar{y}-y\bar{x})=\langle x\bar{y}\rangle_-\;.
\ee
The wedge product corresponds to a so-called biparavector, which can be
used for the description of the electromagnetic field or
the relativistic angular momentum (see also Baylis \cite{Bay99}).
A special notation has been introduced indicating the symmetric and anti-symmetric
contributions of a geometric product.
\be
x\bar{y}=\langle x\bar{y}\rangle_++\langle x\bar{y}\rangle_-\;.
\ee

The basis elements of the $\bar{\bfa{C}}_{3,0}$ paravector algebra can be considered
as the basis vectors of the relativistic vector space. 
These basis elements form a non-cartesian orthogonal basis with respect to the
scalar product defined in Eq.~(\ref{scalar}) 
\be
 e_{\mu}\cdot e_{\nu}
=g_{\mu\nu}\;,
\ee
where $g_{\mu\nu}$ is the metric tensor of the Minkowski space.

The group $SU(2,\bfa{H})$ corresponds to the spin group of the hyperbolic algebra
and its elements can
be used to express rotations and boosts of the
paravectors.
The rotation of a paravector can be expressed as
\be
\label{rota}
x\rightarrow x^\prime=R x\, R^\dagger\;,
\ee
For the boosts one finds the transformation rule
\be
\label{boost}
x\rightarrow x^\prime= B x B^\dagger\;.
\ee
Rotations and boosts are given as
\be
\label{rotmat}
R=\exp{(-i\theta^i\sigma_i/2)}\;,\hspace{0.5cm}B=\exp{(j\xi^i\sigma_i/2)}\;.
\ee
The infinitesimal generators of a Lorentz
transformation can be identified as
\be
\label{gener}
J_i=\sigma_i/2\;,\hspace{0.5cm}K_i= ij\sigma_i/2\;. 
\ee
The generators satisfy the  
Lie algebra of the Lorentz
group.

Boosts are invariant under reversion
$B^\dagger = B$, whereas the conjugated
boost corresponds to the inverse
$\bar{B}=B^{-1}$. For rotations reversion and conjugation correspond both to the inverse
$R^\dagger=\bar{R}=R^{-1}$. The effect of conjugation, reversion, and graduation on the
used hypercomplex units is displayed in Table \ref{invo}.
\begin{table}
\begin{center}
\begin{tabular}{|c|c|c|c|}
\hline
$a$ & $\bar{a}$ & $a^\dagger$ & $\hat{a}$ \\
\hline
$e_0$ & $+$ & $+$ & $+$\\
\hline
$e_i$ & $-$ & $+$ & $-$\\
\hline
$\sigma_i$ & $+$ & $+$ & $+$\\
\hline
$i$ & $-$ & $-$ & $+$\\
\hline
$j$ & $-$ & $+$ & $-$\\
\hline
\end{tabular}
\end{center}
\caption{Effect of conjugation, reversion, and graduation on the used hypercomplex units.\label{invo}}
\end{table}
Note, that graduation is an involution, which 
does not reverse the ordering in a product, i.e., $\widehat{ab}=\hat{a}\hat{b}$.
Conjugation, reversion, and graduation are related by $\bar{a}=\hat{a}^\dagger$.

This was a brief summary of the most important facts. A more detailed representation of
the hyperbolic algebra can be found in \cite{Ulr05}.

\section{Spinors in the hyperbolic algebra}
\label{spinors}
The relationship between relativistic spinors and vectors can be derived
in the same way as in non-relativistic physics. Starting from a parameterization
of a normalized Minkowski vector in terms of spherical coordinates, which
can be generated from a standard vector by a vector transformation, the equivalent
object in spinor space is derived from the corresponding spin transformation.
A normalized spacelike vector $x^\mu$ can be represented as
\be
\label{para}
x^{\mu}=\left(\begin{array}{c}
x^{0}\\
x^{1}\\
x^{2}\\
x^{3}\\
\end{array}\right)=
\left(\begin{array}{c}
\sinh{\xi}\\
\cosh{\xi}\,\sin{\theta}\,\cos{\phi}\\
\cosh{\xi}\,\sin{\theta}\,\sin{\phi}\\
\cosh{\xi}\,\cos{\theta}\\
\end{array}\right)\;.
\ee
In the limit of $\xi\rightarrow 0$ the
vector reduces to a non-relativistic vector in spherical
coordinates. 
The vector can be obtained 
from a standard vector $x^\mu=(0,0,0,1)$ with a
Lorentz transformation of the form
\be
\label{lform}
L= e^{-i\phi J_3}\,e^{-i\theta J_2}\,e^{-i\xi K_3}\;. 
\ee
With the generators of Eq.~(\ref{gener}) the corresponding spin transformation can be written
as 
\be
\label{sform}
S= e^{-i\phi \sigma_3/2}\,e^{-i\theta \sigma_2/2}\,e^{j\xi \sigma_3/2}\;. 
\ee
The spinor, corresponding to the above vector, is obtained in the conventional picture from a multiplication
of the two-component standard spinor $\chi^i=(1,0)$ by the above spin transformation
\be
\label{compspin}
\psi^i=S\chi^i\;.
\ee
The elements of the Pauli algebra are represented here as $2\times 2$ matrices.
The hyperbolic spinor has two components $\psi^i\in \bar{\bfa{H}}^2$. The
bar symbol indicates that the correlation, which maps the elements of the spinor to its
dual space, is defined with conjugation as given in Eq.~(\ref{conj}). With the relation
$\bar{S}=S^{-1}$ it is easy to show that the spinor is normalized
\be
\bar{\psi}_i\psi^i=1\;.
\ee

This representation provides a consistent framework for relativistic calculations \cite{Ulr05}.
However,  this picture requires an explicit matrix representation of the algebra. 
Hestenes \cite{Hes66} suggested to identify the spinor directly
with the spinor transformation itself to obtain a spinor in a pure algebraic form.
This concept can be adopted also in the
current context. The Clifford algebraic spinor is therefore defined as
\be
\label{algspin}
\psi = S\;.
\ee

The spinor can
be expanded into a component structure with an even number of basis vectors
\be
\label{exalgspinor}
\psi = \psi^0+\frac{\psi^{\mu\nu}}{2!}\langle  e_\mu\bar{e}_\nu\rangle_-
+\frac{\psi^{\mu\nu\sigma\rho}}{4!}\langle e_\mu\bar{e}_\nu e_\sigma\bar{e}_\rho\rangle_-\;,
\ee
where only the anti-symmetric contributions of the algebra products are considered.
Note, that this structure is not an element of the even Clifford algebra.
It is a mixture of even and odd elements with respect to the grade involution.
The reason lies in the paravector algebra and the element $e_0=1$, which is
invariant under the grade involution. Baylis \cite{Bay99} introduces the
terminology of a paravector grade. However, it is shown below that a spinor
can not be identified uniquely as an element of even paravector grade.

The factors in Eq.~(\ref{exalgspinor}) are introduced by convention. They indicate that not all elements
in this expansion are linear independent. The spinor consists
of a scalar part, six independent components of a biparavector, and
one pseudoscalar contribution. The last term could therefore be 
expressed also in the simplified form
\be
ij\eta=\frac{\psi^{\mu\nu\sigma\rho}}{4!}\langle e_\mu\bar{e}_\nu e_\sigma\bar{e}_\rho\rangle_-\;,
\ee
with the pseudoscalar $ij\eta$.
The explicit form of the spinor components for the parametrization of
Eq.~(\ref{sform}) is given in the appendix.
The eight independent components are included also in the two-component spinor of
Eq.~(\ref{compspin}). Explicitly one finds
\be
\label{excompspin}
\psi^i=
\left(\begin{array}{c}
\psi^{0}+i\psi^{21}+j\psi^{30}+ij\psi^{0123}\\
\psi^{31}+i\psi^{32}+j\psi^{10}+ij\psi^{20}\\
\end{array}\right)\;.
\ee

The only non-trivial operators that can be generated by the basis elements of the hyperbolic algebra are
$ij$, $i\sigma_i$, and $j\sigma_i$. Together with the unity they form the subalgebra $\bfa{R}_{3,0}$,
which will be denoted here as spinor algebra.
The effect of these operators on the spinor has to be investigated to 
proof the one to one relationship between the spinors given in Eqs.~(\ref{compspin}) and (\ref{algspin}).
For the two-component structure the elements of the Pauli algebra have to be replaced by their
explicit $2\times 2$ matrix representation. If the algebraic spinor is
represented in the form
\bea
\label{exalgspin}
\psi&=&\psi^{0}+\psi^{32}i\sigma_1+\psi^{13}i\sigma_2+\psi^{21}i\sigma_3 \\
&&+\psi^{10}j\sigma_1+\psi^{20}j\sigma_2+\psi^{30}j\sigma_3+ij\psi^{0123}\;,\nonumber
\eea
the proof of this one to one correspondence is straightforward.

From Eq.~(\ref{exalgspin}) it is obvious that the spinor can be expanded also into an odd number of basis vectors
\be
\label{algspinor}
\psi = \psi^{\mu}\, e_\mu+\frac{\psi^{\mu\nu\sigma}}{3!} \langle e_\mu\bar{e}_\nu e_\sigma\rangle_-\;,
\ee
which simply leads to a relabelling of the spinor components.
This spinor is formed by a paravector and a triparavector. The triparavector
is calculated as \cite{Bay99}
\bea
\langle e_\mu\bar{e}_\nu e_\sigma\rangle_-&=&\frac{1}{3!}(e_\mu\bar{e}_\nu e_\sigma+e_\nu \bar{e}_\sigma e_\mu
+e_\sigma \bar{e}_\mu e_\nu \\
&&-e_\nu \bar{e}_\mu e_\sigma-e_\mu \bar{e}_\sigma e_\nu -e_\sigma\bar{e}_\nu e_\mu )\nonumber\;.
\eea

The divisor in Eq.~(\ref{algspinor}) indicates again that
not all elements of the triparavector are linear independent. In fact, there are only
four independent components. One could therefore also
write the last term in the simplified form
\be
ij\eta^{\mu}\,e_\mu =\frac{\psi^{\mu\nu\sigma}}{3!}\langle e_\mu\bar{e}_\nu e_\sigma\rangle_-\;,
\ee
where $ij\eta^{\mu}$ has the structure of a pseudovector.
Though this representation
looks like a sum of a vector and a pseudovector, keep in mind that in the terminology of Clifford
algebras the spinor is considered as an element of a
minimal left ideal. It is only multiplied from the left with other elements of the
algebra, whereas a vector transforms according to Eqs.~(\ref{rota}) and (\ref{boost}).

\section{Spinor product}
In the conventional picture the scalar product of two spinors is
based on the correlation, which maps the elements of the spinor space to
their dual space. If the spinor corresponds to a two-component column
vector the correlation is represented with transposition and conjugation. One can therefore write
\be
\label{convscalar}
\varphi\circ \psi=\bar{\varphi}_i\psi^i\;,
\ee
with $\varphi,\psi\in \bar{\bfa{H}}^2$. The hermitian product will be denoted in the following as spinor product.

Based on Eq.~(\ref{scalar}) the spinor product of two algebraic spinors 
can be defined as
\be
\label{algscalar}
\varphi\circ \psi=\varphi \cdot\psi+j\varphi\cdot\psi e_3\;.
\ee
The second term corresponds to a projection in the direction of the $z$-axis.
An explicit calculation using Eqs.~(\ref{excompspin}) and (\ref{exalgspin})
shows that the above spinor products for the column spinor and the algebraic spinor
are equivalent.
The square of the spinor product can be calculated in terms of relative coordinates as
\be
\label{ampl}
\vert\varphi\circ \psi\vert^2=\cos^2({\theta/2})(1+ij\sinh{\xi}\sin{\phi})\;,
\ee
where $\psi$ has been chosen to be in its standard frame.
Note, that the same expression can be derived also in
momentum space.

This square appears in physics in the calculation of cross sections, for example in the scattering
of polarized electrons by a spinless nucleus like oxygen $^{16}O$ (see, e.g., Perkins \cite{Per87}). The spinor $\psi$
can be chosen to represent the spin structure of the incoming electron beam 
with momentum and polarization $(m_s=+1/2)$ aligned in the
direction of the z-axis. The scattered electron beam, still polarized in a $m_s=+1/2$
state, corresponds to $\varphi$. 
In the case of elastic scattering Eq.~(\ref{ampl}) reduces to the factor $\cos^2({\theta/2})$, which is
equal to the contribution of the electron spin to the Mott formula in the conventional
mathematical formulation of the problem. For inelastic scattering the second term in 
Eq.~(\ref{ampl}) appears, which is proportional to the pseudoscalar $ij$ of the hyperbolic algebra.

The square of the spinor product is a factor in the cross section of the process,
i.e., the number of electrons counted with a certain direction, energy, and momentum is directly related to Eq.~(\ref{ampl}). Since this 
number is
clearly a real number, one may ask whether the hyperbolic complex part of Eq.~(\ref{ampl}) is of physical
relevance.

\section{Summary}
Spinors can be represented in an algebraic form within the
three-dimensional complex Clifford algebra $\bar{\bfa{C}}_{3,0}$.
The conventional two-component 
hyperbolic spinor is equivalent to an expansion of even and odd elements of the Clifford algebra.
This is in contrast to the common understanding
of a spinor as an element of the even Clifford algebra. The reason for this is given by the fact 
that a paravector algebra is used instead of a vector algebra.

The algebraic spinor must not be multiplied from the right by an element of
the Pauli algebra, if the mass operator \cite{Ulr05}, which can be represented in terms
of the spinor algebra $\bfa{R}_{3,0}$, is acting on it.
This is required in the $\bfa{R}_{1,3}$ algebra in order to
keep the spinor within the even algebra, when it is multiplied by the
odd grade Dirac operator.

The scalar product of the Clifford algebra can be used to define a spinor product,
which is equivalent to the conventional scalar product of a column spinor.

\appendix
\section{Algebraic spinors}
The components of the Clifford algebraic spinor are listed below.
They correspond to the parametrization given in
Eq.~(\ref{sform}).
\bea
\psi^{0}&=&\cos{\phi/2}\cos{\theta/2}\cosh{\xi/2}\nonumber\\
\psi^{10}&=&\cos{\phi/2}\sin{\theta/2}\sinh{\xi/2}\nonumber\\
\psi^{20}&=&\sin{\phi/2}\sin{\theta/2}\sinh{\xi/2}\nonumber\\
\psi^{30}&=&\cos{\phi/2}\cos{\theta/2}\sinh{\xi/2}\nonumber\\
\psi^{12}&=&\sin{\phi/2}\cos{\theta/2}\cosh{\xi/2}\nonumber\\
\psi^{31}&=&\cos{\phi/2}\sin{\theta/2}\cosh{\xi/2}\nonumber\\
\psi^{32}&=&\sin{\phi/2}\sin{\theta/2}\cosh{\xi/2}\nonumber\\
\psi^{3210}&=&\sin{\phi/2}\cos{\theta/2}\sinh{\xi/2}
\eea
Note, that the components of the spinor are anti-symmetric with respect
to their indices.
The ordering in the indices has been chosen to give positive values for all
components. The elements appear partly with reversed indices in the
explicit form of the spinors in Eqs.~(\ref{excompspin}) and (\ref{exalgspin}) .

It is an interesting point that in the non-relativistic limit $\xi\rightarrow 0$ a vector can be formed that 
corresponds to a parametrization of a rotation with a $4\pi$
symmetry.
\be
x^i=\left(\begin{array}{c}
\psi^{32}\\
\psi^{13}\\
\psi^{21}
\end{array}\right)=
\left(\begin{array}{c}
\sin{\phi/2}\sin{\theta/2}\\
-\cos{\phi/2}\sin{\theta/2}\\
-\sin{\phi/2}\cos{\theta/2}
\end{array}\right)\;.
\ee

\end{document}